\documentclass[prb,showpacs,showkeys,superscriptaddress,reprint]{revtex4-1}
\usepackage{amsmath}
\usepackage{amsfonts}
\usepackage{amssymb}
\usepackage{hyperref}
\usepackage{graphicx}
\usepackage{cleveref}
\usepackage[none]{hyphenat}
\usepackage{mathptmx}
\begin{document}
\title{Improved superconducting properties of skutterudite $\text{La}_3\text{Co}_4\text{Sn}_{13}$\\ with indium substitution }
\author{P. Neha}
\affiliation{School of Physical Sciences, Jawaharlal Nehru University, New Delhi 110067,India}
\author{P. Srivastava}
\affiliation{School of Physical Sciences, Jawaharlal Nehru University, New Delhi 110067,India}
\author{R. Jha}
\affiliation{School of Physical Sciences, Jawaharlal Nehru University, New Delhi 110067,India}
\affiliation{National Physical Laboratory, New Delhi 110012, India}
\author{Shruti}
\affiliation{School of Physical Sciences, Jawaharlal Nehru University, New Delhi 110067,India}
\author{V.P.S. Awana}
\affiliation{National Physical Laboratory, New Delhi 110012, India}
\author{S. Patnaik}
\email{spatnaik@mail.jnu.ac.in}
\affiliation{School of Physical Sciences, Jawaharlal Nehru University, New Delhi 110067,India}
\begin{abstract}
We report on two fold increase in superconducting transition temperature of $\text{La}_3\text{Co}_4\text{Sn}_{13}$ by substituting indium at the tin site. The transition temperature of this skutterudite is observed to increase from 2.5 K to 5.1 K for 10\% indium substituted sample. The band structure and density of states calculations also indicate such a possibility. The compounds exhibit type - II superconductivity and the values of lower critical field ($\text{H}_{\text{c1}}$), upper critical field ($\text{H}_{\text{c2}}$), Ginzburg-Landau coherence length ($\xi$), penetration depth ($\lambda$) and GL parameter ($\kappa$) are estimated to be 0.0028 T, 0.68 T, 21.6 nm, 33.2 nm and 1.53 respectively for $\text{La}_3\text{Co}_4\text{Sn}_{11.7}\text{In}_{1.3}$. Hydrostatic external pressure leads to decrease in transition temperature and the calculated pressure coefficient is $\sim $ -0.311 K/GPa . Flux pinning and vortex activation energies also improved with indium addition. Only positive frequencies are observed in phonon dispersion curve that relate to the absence of charge density wave or structural instability in the normal state.
\end{abstract}
\email[corresponding author:]{spatnaik@mail.jnu.ac.in}

\maketitle
\section*{Introduction}
Subsequent to the pioneering work by Ramaika et al.\cite{ref1}, the ternary intermetallic stannides  with general formula $\text{R}_3\text{M}_4\text{Sn}_{13}$ have attracted considerable attention in the recent past.  Here R is an alkaline or rare earth metal and M is a transition metal. There is renewed interest in these compounds  because of their manifold properties such as occurrence of charge density wave (CDW), coexistence of superconductivity and magnetism, thermoelectric properties, and Kondo lattice etc[1-19]. In particular, the compound $\text{La}_3\text{Co}_4\text{Sn}_{13}$ crystallizes in a cubic structure (space group Pm-3n) which is analogous to $\text{Yb}_3\text{Rh}_4\text{Sn}_{13}$ and belongs to the skutterudite sub-family. The exciting properties of skutterudites, as thermoelectric materials and towards realization of unconventional superconductivity, rest with their peculiar lattice arrangement. As depicted in inset of Figure\ref{fig1}, the basic structural unit is a polyhedral cage with the transition metal at the centre. When the cage is large compared to trapped atoms, the anharmonic phonon modes of transition metal couples with electrons of the cage leading to enhanced electron - phonon correlation.  $\text{La}_3\text{Co}_4\text{Sn}_{13}$ has been found to be a superconductor with a transition temperature ($ \text{T}_\text{c} $) of about 2.5 K along with coexisting CDW and structural distortion \cite{ref1,ref4,ref8}. Such intermetallic superconductors are extremely susceptible to lattice instabilities and a recent high pressure study on $\text{La}_3\text{Co}_4\text{Sn}_{13}$ has confirmed a rare instance of increase in $ \text{T}_\text{c} $ with pressure \cite{ref4}.  Moreover several other $\text{R}_3\text{M}_4\text{Sn}_{13}$ compounds exhibit higher superconducting transition. The examples include $\text{Ca}_3\text{Rh}_4\text{Sn}_{13}$ (8.7 K), $\text{Yb}_3\text{Rh}_4\text{Sn}_{13}$ (5 K) and $\text{Ca}_3\text{Ir}_4\text{Sn}_{13}$ $\left(7 \text{K}\right)$. In fact, smaller Ca ion results in $\text{T}_\text{c}$ higher than $\text{Sr}_3\text{Ir}_4\text{Sn}_{13}$ which is reflective of increased $ \text{T}_\text{c} $ under chemical pressure \cite{ref6}. In this communication, we report over two fold increase in superconducting transition temperature in $\text{La}_3\text{Co}_4\text{Sn}_{13}$ by partial substitution of indium in place of tin. We note that atomic radius of indium is larger than tin and we provide a detailed investigation of band structure on superconducting properties of indium doped $\text{La}_3\text{Co}_4\text{Sn}_{13}$. The superconducting transition temperature of $\text{La}_3\text{Co}_4\text{Sn}_{13-\text{x}}\text{In}_{\text{x}}$  increased from 2.5 K for undoped $\text{La}_3\text{Co}_4\text{Sn}_{13}$ to 5.1 K with 10\% indium substitution. In effect, our results establish the possibility of fine tuning of lattice instabilities and density of states at Fermi level in caged skutterudite $\text{La}_3\text{Co}_4\text{Sn}_{13}$ by chemical substitution towards achieving higher $\text{T}_{\text{c}}$ phases. 

\section*{Experiment}
The stannide samples were synthesized by melting metals in a single-arc furnace (Miller). A series of $\text{La}_3\text{Co}_4\text{Sn}_{13-\text{x}}\text{In}_{\text{x}}$ (where x = 0, 0.65 and 1.3 corresponding to 0\%, 5\% and 10\% indium substitution) were synthesized in two steps. All the constituent elements, La(99.99\%), In(99.999\%), Sn(99.999\%), Co(99.999\%), were taken in proper stoichiometric ratio and melted on the water cooled copper hearth by arc melting in argon atmosphere. The melted ingots were ground, palletized and then sealed in quartz ampoules in $ 10^{-3} $ mbar vacuum and annealed at $870^\circ \text{C}$ for 12 days. Room-temperature X-ray diffraction measurement was done with Cu-K$\alpha $ radiation by RIGAKU powder X-ray Diffractometer (Miniflex-600). All the samples were identified within the expected cubic $\text{Yb}_3\text{Rh}_4\text{Sn}_{13}$-type phase (space group Pm-3n). EDAX and SEM data were obtained from Bruker AXS Micro-analyser and from Zeiss EVO40 SEM analyzer respectively. Several points were selected and averaged to get the actual atomic percentage of the samples. The resistivity was measured using linear four probe technique using copper wire and silver epoxy. Magnetization measurements were carried out using 14 Tesla $\textit{Cryogenic}$  Physical Property Measurement System (PPMS). Pressure-dependent resistivity measurements were performed in the Physical Property Measurements System (PPMS-14T, Quantum Design) using an HPC-33 Piston type pressure cell with Quantum Design DC resistivity option. 

\section*{Results and Discussion}
\begin{figure}[h!]
\centering
\includegraphics[scale=0.55]{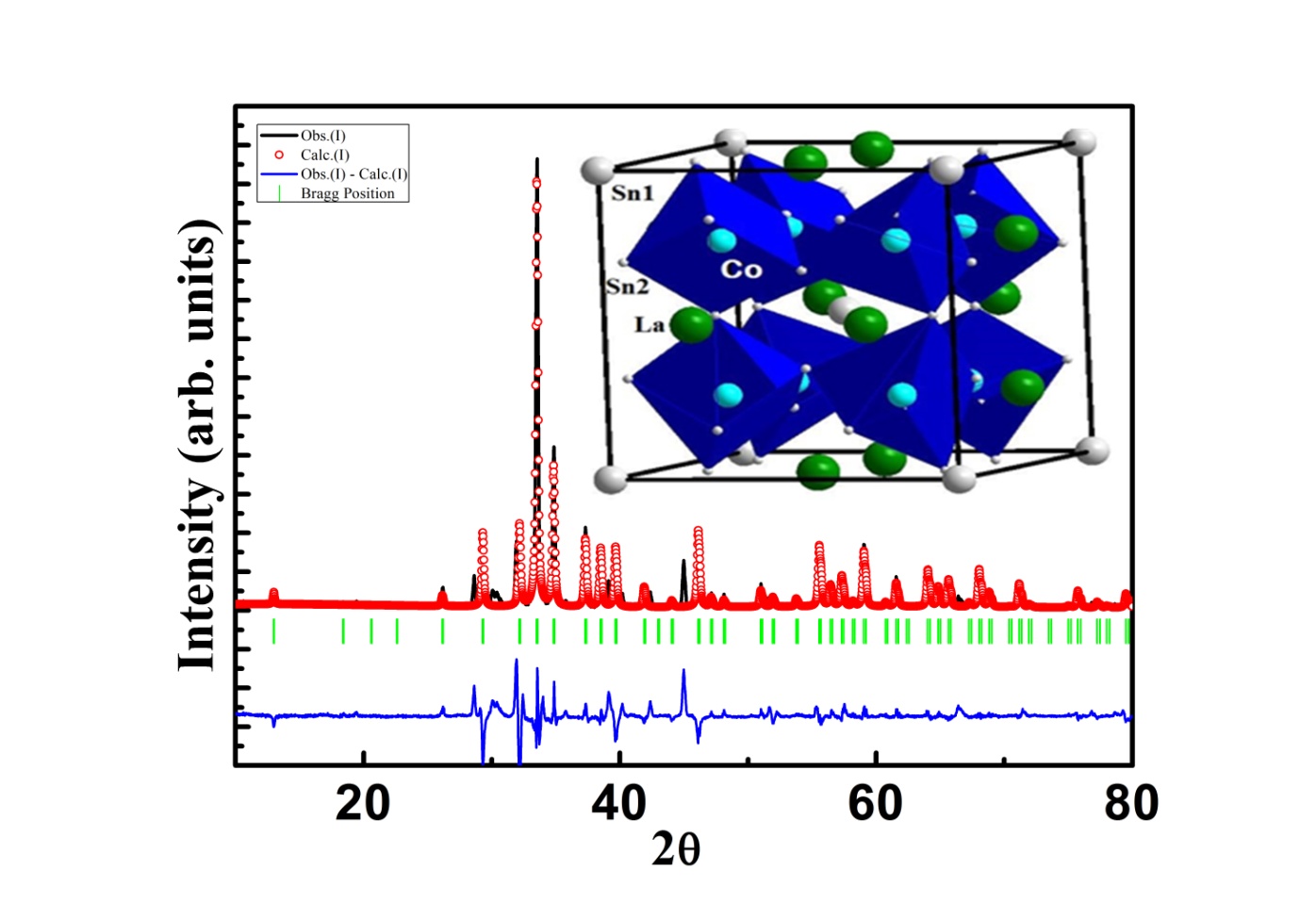}
\caption{\label{fig1}Reitveld refinement of XRD data of $\text{La}_3\text{Co}_4\text{Sn}_{13}$. Inset shows schematic crystal structure of $\text{La}_3\text{Co}_4\text{Sn}_{13}$.}
\end{figure}

\begin{figure}[h!]
\centering
\includegraphics[scale=0.55]{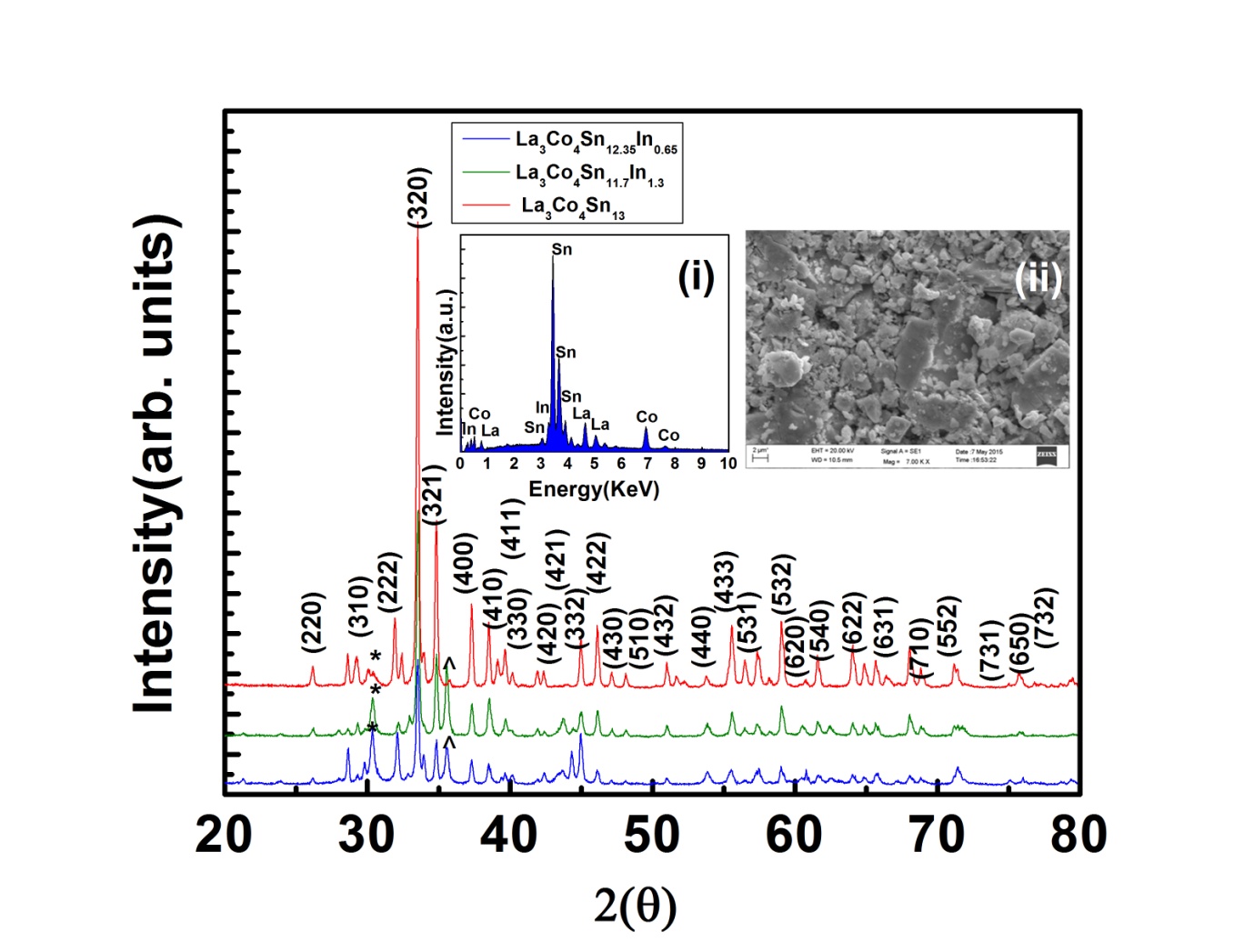}
\caption{\label{fig2}Combined XRD of $\text{La}_3\text{Co}_4\text{Sn}_{13}$, $\text{La}_3\text{Co}_4\text{Sn}_{12.35}\text{In}_{0.65}$ , and $\text{La}_3\text{Co}_4\text{Sn}_{11.7}\text{In}_{1.3}$ samples. Inset (i) shows EDAX data for $\text{La}_3\text{Co}_4\text{Sn}_{11.7}\text{In}_{1.3}$  and inset (ii) depicts SEM image of $\text{La}_3\text{Co}_4\text{Sn}_{11.7}\text{In}_{1.3}$.}
\end{figure}

The main panel of Figure\ref{fig1} shows the Rietveld refined XRD patterns for $\text{La}_3\text{Co}_4\text{Sn}_{13}$.  The inset shows schematic digram of crystal structure.  Evidently, there are two positions for Sn atoms; those that are at the corners of unit cell (Sn1) and the rest that form the polyhedra enclosing the Co atom at the center (Sn2).  Thus Co and Sn2 atoms form corner sharing $ \text{Co}(\text{Sn})_ 6$ trigonal prisms.
  Figure\ref{fig2} compares the XRD patterns for $\text{La}_3\text{Co}_4\text{Sn}_{13-\text{x}}\text{In}_{\text{x}}$ (where x = 0, 0.65 and 1.3)­.The peaks are indexed with $\text{Yb}_3\text{Rh}_4\text{Sn}_{13}$-type structure of space group Pm-3n. The lattice parameter of $\text{La}_3\text{Co}_4\text{Sn}_{13-\text{x}}\text{In}_{\text{x}}$ are estimated to be a = 9.636 $\AA$, 9.637 $\AA$  and  9.639 $\AA$  for  x = 0, 0.65 and 1.3 respectively. Two impurity phases, $\text{La}_3\text{Co}_2$ and $\text{La}_3\text{Co}_{0.05}\text{In}_{0.95}$, have also been identified which are marked by $*$ and $\wedge$  respectively. The EDAX measurements provide the percentage of the constituent elements in the compound. 
For 5\% indium substituted compound, the atomic percentage for La, Co, Sn and In are 14.09\%,  20.74\% and 62.0\% and 3.16\% respectively, while for 10\% indium substituted sample the percentages are 15.1\%, 20.21\%, 58.32\% and 6.37\% respectively (inset (i)). SEM image of $\text{La}_3\text{Co}_4\text{Sn}_{11.7}\text{In}_{1.3}$ is shown in inset (ii) of Figure\ref{fig2}.  It shows grains with varying sizes and the features are unlike layered structures observed in cuprate and chalcogenide superconductors.
 
\begin{figure}[h!]
\centering
\includegraphics[scale=1.5]{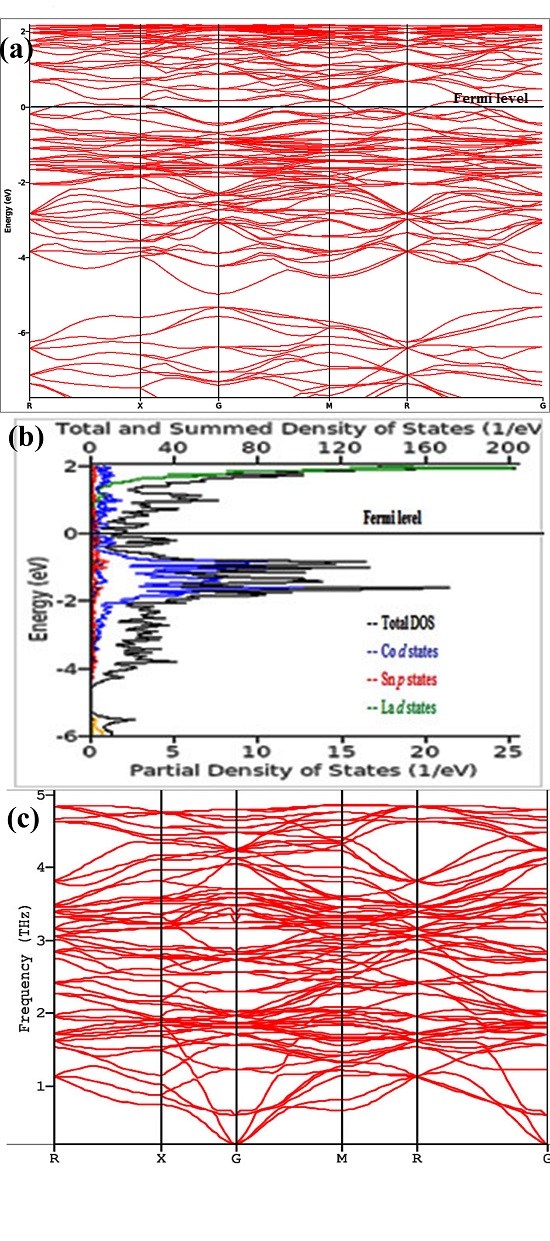}
\caption{\label{fig3}(a) Band structure of $\text{La}_3\text{Co}_4\text{Sn}_{13}$. (b) Calculated density of states, and (c) Phonon dispersion curve of $\text{La}_3\text{Co}_4\text{Sn}_{13}$.}
\end{figure}

In Figure\ref{fig3} we discuss the electronic density of states analysis of $\text{La}_3\text{Co}_4\text{Sn}_{13}$. Using VASP 5.3 codes we have studied the electronic band structure of $\text{La}_3\text{Co}_4\text{Sn}_{13}$ and its possible tunability towards achieving higher $ \text{T}_{c}$ phases by chemical substitution. For the calculation of band structure and density of state of the parent compound, we employed PAW (projected augmented waves) and plane wave basis set of 360 eV cut off. The DFT calculations were performed using GGA-PBE (Generalized Gradient Approximation-Perdew­Burke Ernzerhof) approximations. Typically $13 \times 13 \times 13 $ mesh points were used with K-point grid of $0.05 \times 0.05 \times 0.05$ to achieve optimal convergence. 

Figure\ref{fig3}(a) shows results of band structure calculation in high symmetry directions of $ \text{La}_3\text{Co}_4\text{Sn}_{13} $. From the large dispersion of bands near Fermi level it is evident that that the parent compound is metallic in nature. The density of state at Fermi level (Figure\ref{fig3}(b)) reveals that there is strong hybridization of Co 3d orbital and Sn 5p orbital that contributes to observed superconductivity. At the Fermi level the value of DOS is around 15 states/eV [in Formula Unit], which reconfirm metallicity in the parent compound. Presumably, superconductivity arises due to orbital hybridization of Sn2 and Co atoms. The outer Sn1 atoms at the corners of the unit cell, do not contribute to superconductivity. In effect, the tuning of superconducting properties of this skutterudite would depend on effective modification of the density of states at Fermi level by appropriate substitution at the Sn2 site. Further, as seen in Figure\ref{fig3}(c), only positive frequencies are seen in phonon dispersion curve indicating absence of structural instability. This is in contrast to what was reported from specific heat measurements by Liu et al.\cite{ref8}. In the skutterudite family, structural instabilities were also reported in $\text{Sr}_3\text{Ir}_4\text{Sn}_{13}$ and $\text{Ca}_3\text{Ir}_4\text{Sn}_{13}$ where negative frequencies in phonon dispersion curves were observed.  This was associated with CDW transitions around 147 K and 35 K respectively \cite{ref3,ref5,ref9,ref14}.  But we find no evidence for such instabilities either in the parent or in substituted $\text{La}_3\text{Co}_4\text{Sn}_{13}$. \\
\begin{figure}[h!]
\centering
\includegraphics[scale=0.6]{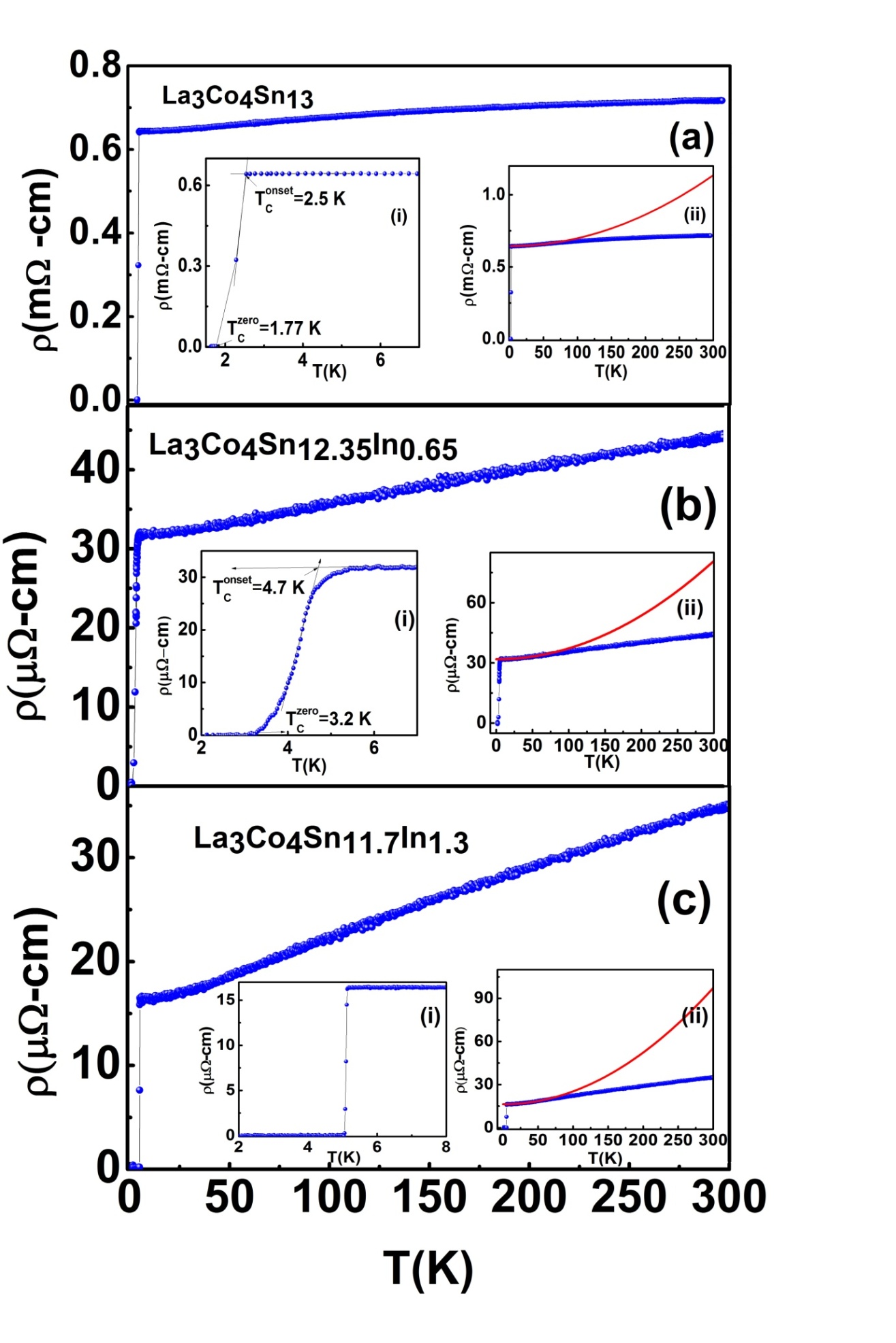}
\caption{\label{fig4}Electrical resistivity is plotted as a function of temperature for (a) $\text{La}_3\text{Co}_4\text{Sn}_{13}$, (b) $\text{La}_3\text{Co}_4\text{Sn}_{12.35}\text{In}_{0.65}$ and (c) $\text{La}_3\text{Co}_4\text{Sn}_{11.7}\text{In}_{1.3}$.  Insets (i) and (ii) in each panel show superconducting transition and  Fermi liquid fitting respectively.}
\end{figure}
 Figure\ref{fig4} shows the temperature dependence of resistivity from 1.7 K to 300 K for $\text{La}_3\text{Co}_4\text{Sn}_{13-\text{x}}\text{In}_{\text{x}}$(x = 0, 0.65 and 1.3).  The insets show magnified superconducting transition and Fermi liquid fitting. As seen in inset of Figure\ref{fig4}(a), the parent $\text{La}_3\text{Co}_4\text{Sn}_{13}$ shows $\text{T}_\text{c}^\text{onset}$ = 2.5 K and $\text{T}_\text{c}^\text{zero}$ = 1.77 K. This value is in accordance with earlier reports \cite{ref4}. Most notably, there is significant enhancement in superconducting transition to $\text{T}_\text{c}^\text{onset}$  = 4.7 K and $\text{T}_\text{c}^\text{zero}$= 3.2 K for 5\% indium doped $\left(\text{La}_3\text{Co}_4\text{Sn}_{12.35}\text{In}_{0.65}\right)$ compound.  On careful analysis we find the deviation from normal state resistivity actually begins from 5.3 K (inset (i) Figure \ref{fig4}(b)). This is suggestive of inhomogeneous distribution of grains. More striking is the low temperature resistivity data of $\text{La}_3\text{Co}_4\text{Sn}_{11.7}\text{In}_{1.3}$ (inset (i) Figure\ref{fig4}(c)) where an extremely sharp transition of $\text{T}_\text{c}^\text{onset}$ = 5.1 K is seen with transition width $ \Delta$T $\sim$ 0.1K. This is more than two times than that for the parent compound.  The residual resistivity ratio (RRR) for $\text{La}_3\text{Co}_4\text{Sn}_{13-\text{x}}\text{In}_{\text{x}}$ is found to be 2.18, 1.37 and 2.15 for x = 0, 0.65 and 1.3 respectively.  But we note that unit of resistivity is in $\text{m}\Omega$  $\text{cm}$ for the parent compound in contrast to $\mu \Omega $  $\text{cm}$ for the doped specimen. The resistivity variation with temperature indicates the metallic nature of the parent as well as doped compounds in the normal state. In contrast, iso-structural compounds with Ge at the Sn place, i.e. $\text{Lu}_3\text{Os}_4\text{Ge}_{13-x}$, $\text{Lu}_3\text{Rh}_4\text{Ge}_{13-x}$, $\text{Y}_3\text{Co}_4\text{Ge}_{13-x}$ etc show semiconducting behaviour in normal state \cite{ref17}. The reason for different normal state conduction mechanism is attributed to the atomic displacement parameter (ADP) and compounds with smaller ADP are found to exhibit metallic ground state while those with large ADP have semiconducting ground state \cite{ref17}. Moreover, no evidence for the presence of CDW is observed either in the parent or in the indium doped samples which has been observed in other stannides such as $\text{Sr}_3\text{Ir}_4\text{Sn}_{13}$ and $\text{Ca}_3\text{Ir}_4\text{Sn}_{13}$ \cite{ref9,ref14}. In $\text{Sr}_3\text{Ir}_4\text{Sn}_{13}$ the CDW emerges because of structural transition from simple cubic to a body centered cubic superstructure with doubling of the cell parameter below 147 K. This is confirmed by optical spectroscopy, Hall, specific heat and NMR measurements \cite{ref9,ref14}.  A first-order phase transition has been observed in single-crystal of $\text{La}_3\text{Co}_4\text{Sn}_{13}$ with a marked peak at T $\sim$152 K by specific heat and NMR measurements\cite{ref8}. The observed transition has been related to a structural change from a simple cubic to a body-centred-cubic super-structure with crystallographic cell doubling.  But we do not observe this behaviour in our polycrystalline samples. Further, we did not find any anomaly in the specific heat measurements (not included here) around the reported CDW transition temperature range.
  
The normal state resistivity curves (Figure\ref{fig4}) have also been fitted to the formula $\rho = \rho_\circ + AT^2$. The values of residual resistivity $(\rho_\circ)$ for $\text{La}_3\text{Co}_4{\text{Sn}}_{13-\text{x}}\text{In}_{\text{x}}$ are 0.644 $\text{m}\Omega$ $\text{cm}$, 31.30 $ \mu \Omega$ $\text{cm}$ 16.45 $ \mu \Omega$ $\text{cm}$; and values for A are  5.482 $\times 10^{-6} \text{m}\Omega$ $\text{cm}/\text{K}^{2}$ , 4.542 $\times 10^{-4} \mu \Omega$  $\text{cm}/\text{K}^{2}$ and  7.632 $ \times 10^{-4}\mu \Omega$  $\text{cm}/\text{K}^2$ for x = 0, 0.65 and 1.3 respectively. It is interesting to note that the residual resistivity decreased for higher indium percentage.  Further, magnitude of parameter A, which provides a scale for the degree of electronic correlation, increased reflecting an enhanced electronic correlation for higher indium percentage. Overall, the normal state resistivity data of $\text{La}_3\text{Co}_4\text{Sn}_{13}$ can be categorized as weak Fermi liquid behaviour\cite{ref9}.
\begin{figure}[h!]
\centering
\includegraphics[scale=0.7]{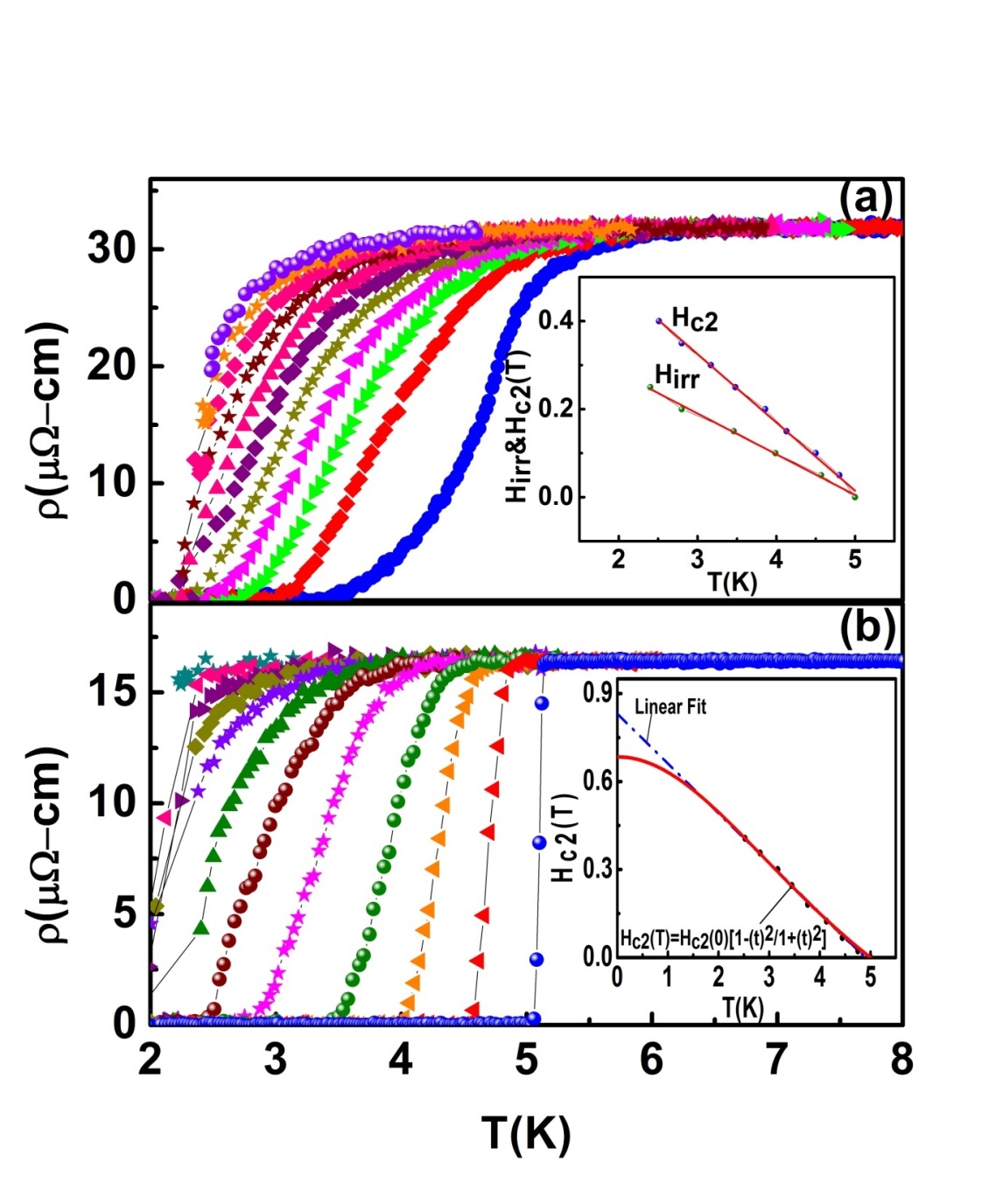}
\caption{\label{fig5}Resistivity variation with respect to temperature of (a) $\text{La}_3\text{Co}_4\text{Sn}_{12.35}\text{In}_{0.65}$ and (b) $\text{La}_3\text{Co}_4\text{Sn}_{11.7}\text{In}_{1.3}$ in the presence of applied magnetic field. For $\text{La}_3\text{Co}_4\text{Sn}_{12.35}\text{In}_{0.65}$ the applied field values are 0 T, 0.1 T, 0.15 T, 0.2 T, 0.25 T, 0.3 T, 0.35 T, 0.4 T, 0.45 T, 0.5 T, and 0.65 T.  Similarly, for $\text{La}_3\text{Co}_4\text{Sn}_{11.7}\text{In}_{1.3}$, the corresponding field values are 0 T. 0.05 T, 0.1 T,0.15 T, 0.2 T, 0.25 T, 0.3 T, 0.35 T, 0.4 T, 0.45 T, 0.55 T and 0.7 T.  Inset (i) shows $\text{H}_\text{c2}$ and $\text{H}_\text{irr}$ variation as a function of temperature for $\text{La}_3\text{Co}_4\text{Sn}_{11.7}\text{In}_{1.3}$.  Inset (ii) shows extrapolated $\text{H}_\text{c2}$ - $\text{T}$ phase diagram according to Ginzburg-Landau model.}
\end{figure}

 Figure\ref{fig5} shows the resistivity variation with temperature at different applied external magnetic field for $\text{La}_3\text{Co}_4{\text{Sn}}_{13-\text{x}}\text{In}_{\text{x}}$ with (a) x= 0.65 and (b) 1.3.  In both cases, the superconducting transition broadens with the application of magnetic fields along with decrease in $\text{T}_\text{c}$. The inhomogeneous nature of x= 0.65 sample is evident in in-field study as well. For x= 1.3, the $\text{T}_\text{c}$(H) values for onset and offset superconducting transition as a function of magnetic field and temperature were ascertained.  The corresponding upper critical $\text{H}_{\text{c2}}$(T) and irreversibility $\text{H}_\text{irr}$(T) fields are plotted in the inset of the Figure\ref{fig5}(a). Using generalized Ginzburg-Landau model for upper critical field $\text{H}_{\text{c2}}$(T) = $\text{H}_{\text{c2}}$(0)[(1- $\text{t}^2$)/ (1+ $\text{t}^2$)] with t = T/$\text{T}_\text{c}$, the H(T) phase diagram is sketched in inset of Figure\ref{fig5}(b). The extrapolated  $\text{H}_\text{c2}$ (0) is 0.68 Tesla.  Using this value, the Ginzburg – Landau coherence length $\xi_{\text{GL}}$= $\left(\Phi_\circ / 2\pi \text{H}_\text{c2}\right)^{1/2}$, with flux quantum $\Phi_\circ=2.07\times 10^{-7}\;\text{G}\;\text{cm}^2$ , is estimated to be  $\xi_\text{GL}(0) \sim   19.5$ nm, 21.6 nm for 5\% , 10\% indium doped specimen respectively. In type - II superconductors the pair breaking occurs mainly in two ways, either by orbital effect or by spin paramagnetic effect. Further, in clean limit the orbital upper critical field is given by $\text{H}_\text{c2}^\text{orbital}$ (0) = -0.72$\text{T}_\text{c}$ [dH/dT]$_{\text{T}=\text{T}_\text{c}}$. This is found to be 0.615 T for 10\% indium substituted sample. On the other hand, using the formula, $\text{H}^{\text{Pauli}}$ =1.82 $\text{T}_\text{c}$ in the weak coupling Pauli paramagnetic limit,$\text{ H}^{\text{Pauli}}$ is found to be 9.1 T.  It is seen that, $\text{H}_\text{c2}^\text{orbital} \;<\; \text{H}_\text{C2} \;(0)\;<\;\text{H}^{\text{Pauli}}$  and this indicates that upper critical field in $\text{La}_3\text{Co}_4{\text{Sn}}_{13-\text{x}}\text{In}_{\text{x}}$ is Pauli limited.
\begin{figure}[h!]
\centering
\includegraphics[scale=0.6]{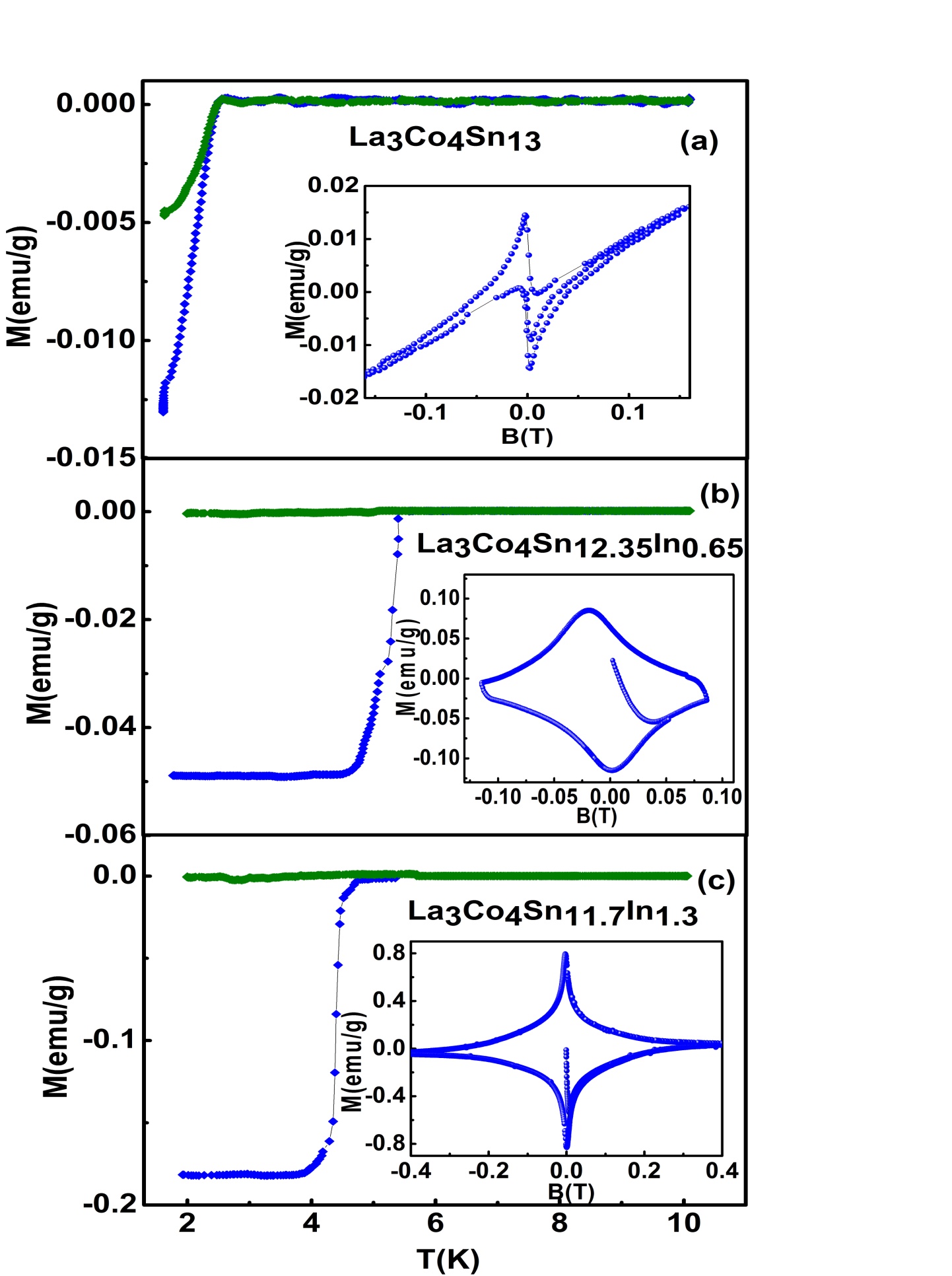}
\caption{\label{fig6}The main panel shows ZFC-FC data at 20 G external field for (a) $\text{La}_3\text{Co}_4\text{Sn}_{13}$, (b) $\text{La}_3\text{Co}_4\text{Sn}_{12.35}\text{In}_{0.65}$, and (c) $\text{La}_3\text{Co}_4\text{Sn}_{11.7}\text{In}_{1.3}$.  The insets show M-H loop at 1.7 K for the corresponding samples.}
\end{figure}

 The magnetization measurements were performed under zero field cooled (ZFC) \& field cooled (FC) protocols. Figure\ref{fig6}  shows the magnetization measurement in ZFC and FC condition in applied field of 20 Gauss. The superconducting transition temperature for $\text{La}_3\text{Co}_4{\text{Sn}}_{13-\text{x}}\text{In}_{\text{x}}$ is marked at 2.2 K, 5.1 K and 4.6 K for x = 0, 0.65 and 1.3 respectively. From the ZFC-FC data of the parent material and the doped compound, we get that the irreversible region to be more in case of indium substituted ones. The increasing irreversibility region indicates that the vortex pinning has been enhanced with indium substitution. The M-H loop taken at 1.7 K is shown in insets of Figure 6. In indium substituted compounds, the M-H loop shows the characteristic features of the type-II superconductors. The lower critical field ($\text{H}_{\text{c1}}$) which is marked by a deviation from linearity in the diamagnetic state is  0.0028T in 10\%  indium doped $\text{La}_3\text{Co}_4\text{Sn}_{13}$. Using the values of $\text{H}_{\text{c2}}$ and $\text{H}_{\text{c1}}$, penetration depth ($\lambda$) and Ginzburg Landau parameter ($\kappa$) are calculated by $\text{H}_{\text{c1}} = \phi_\circ/(2\pi\lambda ^2)$ and $\kappa = \lambda/ \xi$. The values of $\lambda$ = 33.2 nm and $\kappa = 1.53$ are obtained for  $\text{La}_3\text{Co}_4\text{Sn}_{11.7}\text{In}_{1.3}$.
\begin{figure}[h!]
\centering
\includegraphics[scale=0.3]{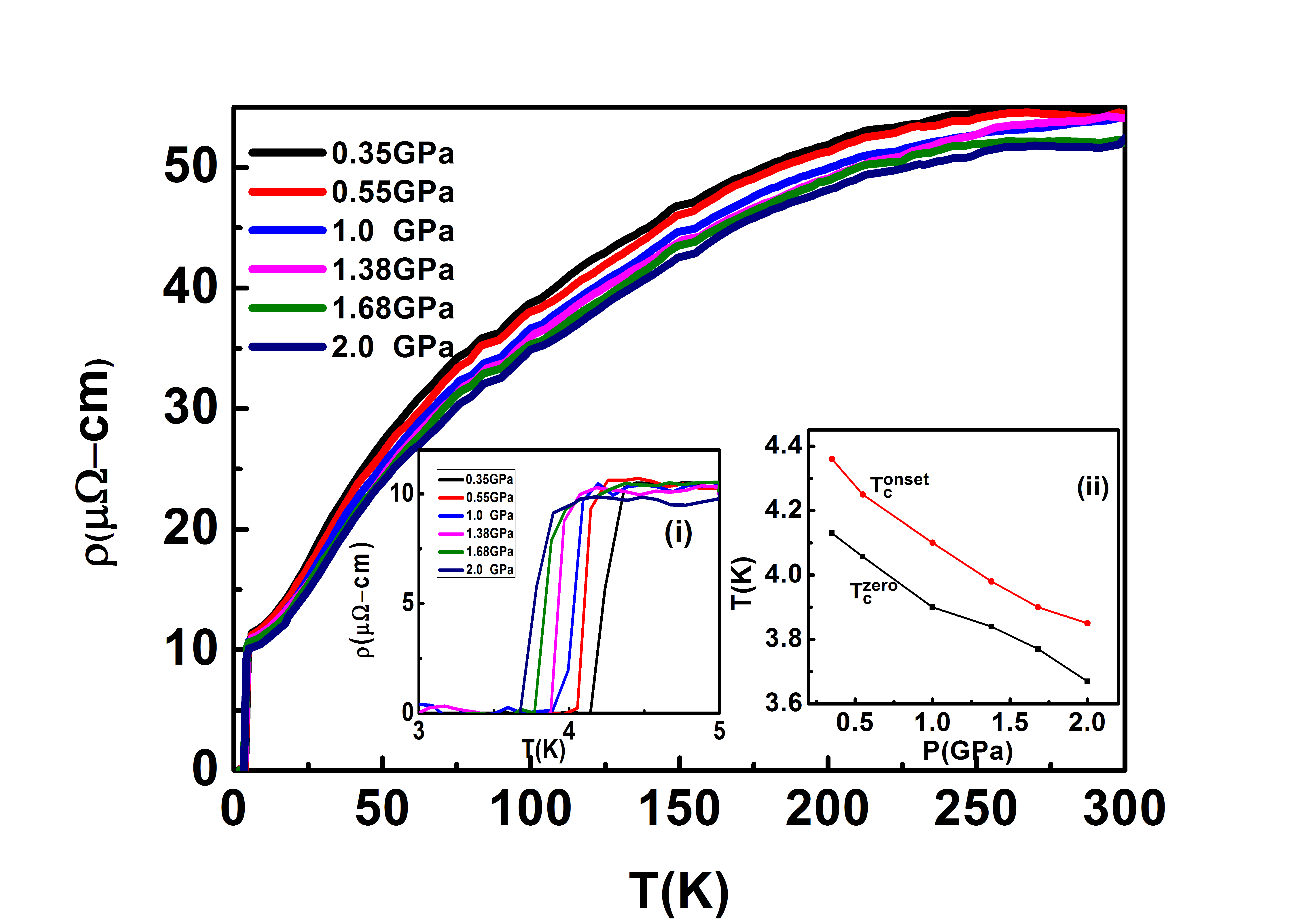}
\caption{\label{fig7}Variation of resistivity of $\text{La}_3\text{Co}_4\text{Sn}_{11.7}\text{In}_{1.3}$  upto room temperature at various external pressure. Superconducting transition of $\text{La}_3\text{Co}_4\text{Sn}_{11.7}\text{In}_{1.3}$ at pressure between 0.35 GPa and 2 GPa is shown in inset(i) . Variation of $\text{T}_{c}^\text{onset} $ and $\text{T}_{c}^\text{zero} $ at pressure upto  2 GPa is plotted in inset(ii)}
\end{figure}

Like magnetic field, applied pressure also tunes various parameter of superconductivity like density of states at Fermi level and  electron - phonon coupling etc. In  conventional superconductors  $\text{T}_\text{c}$ decreases with pressure\cite{ ref21}, but in cuprates  $\text{T}_\text{c}$ first increases till optimum pressure and then decreases on further increase in pressure\cite{ref20}. Figure\ref{fig7} shows the impact of pressure on the $\text{T}_\text{c}$ of 10\% Indium substituted sample. The main panel of Figure\ref{fig7} shows the variation of resistivity till 300 K at different pressure ranging between 0.35 GPa and 2 GPa.  Variation in superconducting transition at different pressure is shown in inset(i) of Figure\ref{fig7} which clearly shows that $\text{T}_\text{c}^\text{onset}$ as well as $\text{T}_{c}^\text{zero}$ are decreasing monotonously. At P= 0.35 GPa, $\text{T}_\text{c}^\text{onset} $ decreases from 5.1 K to 4.3 K  and with increasing pressure till 2 GPa it decreases to 3.85 K. Similarly the $\text{T}_{c}^\text{zero} $ also decreases with pressure from 5 K to 4.13 K for 0.35 GPa and further to 3.67 for 2GPa. The values of corresponding pressure coefficients  ${\text{dT}_\text{c}^\text{onset}}/{\text{dP}}$ and  ${\text{dT}_\text{c}^\text{zero}}/{\text{dP}}$ are  -0.311, -0.267 K/GPa respectively. This result is in contrast with the earlier report on pressure effects in parent compound $\text{La}_3\text{Co}_4\text{Sn}_{13} $ \cite{ref4} where $\text{T}_\text{c}$ increased with application of external pressure. This indicates with indium doping and consequent positive chemical pressure optimal $\text{T}_{c}$ increase has been achieved. Although not reported here, higher indium substitution led to inhomogeneous samples with no clear evidence for increase in transition temperature. We note that $\text{La}_3\text{Co}_4\text{Sn}_{13}$ is assigned as a BCS superconductor and in weak coupling theory, $ \text{T}_\text{c} $ $\sim $ $ \theta_\text{D} $ $ \text{exp}$ $\left( \frac{-1}{\text{N}{\left(\text{E}_{\text{F}}\right){\text{V}_0}}}\right) $, where $ \theta_\text{D} $ is the Debye temperature,$ \text{V}_0 $ is electron phonon coupling strength, and  density of state  is $ {\text{N}{(\text{E}_\text{F})}}$ goes as $ {\text{m}^{*}{\text{n}^{1/3}}} $ where $ \text{m}^{*} $ is effective mass and n is carrier density. In Figure\ref{fig7} along with decrease in $\text{T}_\text{c}^\text{onset}$, we find increase in metallicity with increase in external pressure. This implies we can not assign the decrease in $ \text{T}_\text{c} $ with pressure merely to change  in carrier concentration. Further a low carrier concentration $ \sim $ $ 10^{19} $ per cc is estimated from room temperature Hall measurements. Therefore some aspects of unconventional superconductivity involving effective mass $ \text{m}^{*} $ is indicated in indium substituted optimal $\text{La}_3\text{Co}_4\text{Sn}_{13}$ compositions. 
\begin{figure}[h!]
\centering 
\includegraphics[scale=0.6]{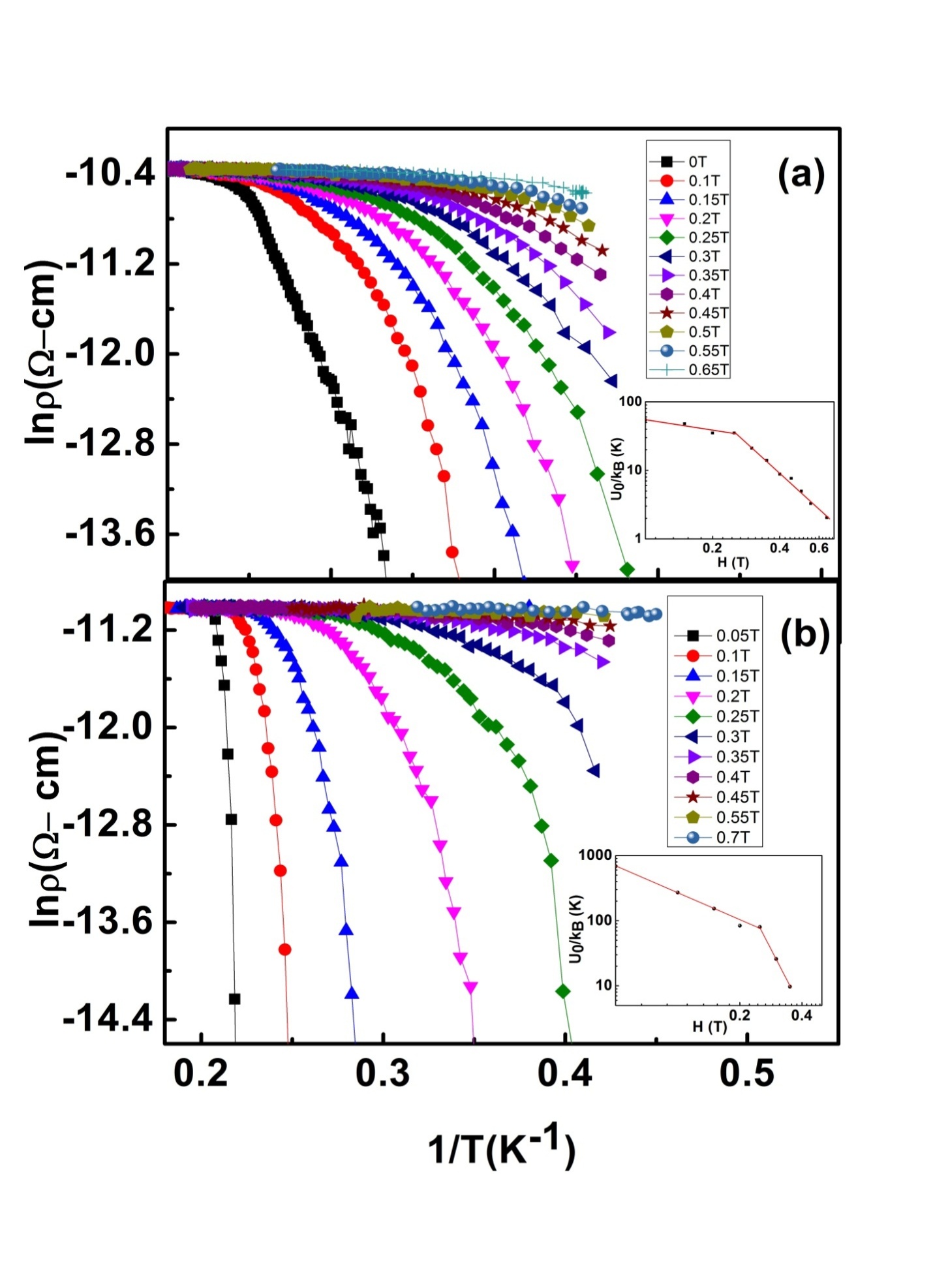}
\caption{\label{fig8}(a) Arrhenius plot of $\text{La}_3\text{Co}_4\text{Sn}_{12.35}\text{In}_{0.65}$. Variation of vortex activation energy as a function magnetic field is plotted in the inset (b) Arrhenius plot of $\text{La}_3\text{Co}_4\text{Sn}_{11.7}\text{In}_{1.3}$. Insets show magnetic field dependence of vortex activation energy.}
\end{figure}

 The broadening in the resistivity curve on application of the applied external magnetic field is due to the vortex motion in the mixed state and dissipation of energy because of the vortex motion. Vortex activation energy variation with applied magnetic fields provides the scale for a material towards potential applications. This Arrhenius plot provides the flux flow activation energy dependence on the applied magnetic field through the relation $\rho (\text{T}, \text{B}) = \rho_\circ \exp (-\text{U}_\circ / \text{k}_\text{B}T)$.  The activation energy $\text{U}_\circ$ is determined by taking the slope of the linear part of Arrhenius plot while $\rho _\circ$ is the field independent pre-exponential factor. The activation energy exhibits different power-law dependences on a magnetic field, i.e. $\text{U}_\circ (B) \propto \text{B}^{-\text{n}}$. The values of $\text{U}_\circ$, are deduced from the limited temperature intervals below $ \text{T}_\text{c}$, in which the data of the Arrhenius plot of $\rho (\text{T})$ yield straight lines. The Arrhenius plots are shown in the Figure\ref{fig8} and activation energy is plotted in the inset. The straight line behaviour over 3 decades of the resistivity data validates the thermally activated flux flow (TAFF) process as described by the Arrhenius law. For 5\% In doped sample (Figure\ref{fig8}(a)), the activation energy dependence on the magnetic field is  $\text{U}_\circ$ (B) $\propto \text{B}^{-0.16}$ for B $ < $ 0.25 T and $\text{U}_\circ$ (B)$ \propto$ $\text{B}^{-0.93}$ for B $>$ 0.25 T.  Here, the activation energy is $\text{U}_\circ /\text{k}_\text{B} \sim  52.8 $ K in low magnetic field (0.01 T) and            $ \text{U}_\circ / \text{k}_\text{B}\sim 2.0$  K in high field region (0.65 T). For 10\% In doping (shown in Figure\ref{fig8}(b)), the values of activation energy are 684.6 K (0.05 T) and 9.4 K (0.35 T).  The exponent n in the power law equation $\text{U}_\circ \text{(B)} \propto \text{B}^\text{n}$ are 0.44 for B $<$ 0.25 T and 1.97 for B $>$ 0.25 T.\\
\section*{Conclusion}
 In summary, the superconducting transition temperature increased more than twice compared to $\text{La}_3\text{Co}_4\text{Sn}_{13}$ (from 2.5 K to 5.1 K) after substituting indium in place of tin. The substituted indium effectively tunes the lattice instabilities of the caged skutterudite leading to a higher $\text{T}_\text{c}$ phase.  Theoretical calculation of band structure and density of states indicate that the Co 3d and Sn 5p orbitals contribute to the higher density of states about the Fermi level. It is understood that added indium would be preferentially substituting Sn2 atoms leading to enhancement in $\text{T}_\text{c}$. Calculated phonon dispersion curves rule out the presence of lattice instability. The parent and doped stannides are found to exhibit weak electronic correlation that increased with increasing indium substitution. The value of lower critical field, upper critical field, Ginzburg-Landau coherence length , penetration depth  and GL parameter  are found to be 0.0028 T, 0.68 T, 21.6 nm, 33.2 nm and 1.53 respectively for the 10\% indium substituted $\text{La}_3\text{Co}_4\text{Sn}_{13}$ sample. The superconducting transition temperature of $\text{La}_3\text{Co}_4\text{Sn}_{11.7}\text{In}_{1.3}$ decreased on application of external pressure in contradiction to what is reported for $\text{La}_3\text{Co}_4\text{Sn}_{13}$.  
\section*{Acknowledgements}
PN acknowledges UGC for support under BSR fellowship program. PS and RJ acknowledge UGC for support from  D. S. Kothari Fellowships. SP thanks the FIST program of the Department of Science and Technology, Government of India for low temperature high magnetic field facilities at JNU. Technical support from AIRF (JNU) is acknowledged.


\end{document}